\pdfoutput=1
\RequirePackage{ifpdf}
\documentclass{JINST} 

\usepackage[utf8]{inputenc}
\usepackage{amsmath}

\graphicspath{{figures/}}

\title{T3B - An Experiment to measure the Time Structure of Hadronic Showers}

\author {Frank Simon\footnote{corresponding author, email: fsimon@mpp.mpg.de}, Christian Soldner, Lars Weuste\\{\it Max-Planck-Institut f\"ur Physik, F\"ohringer Ring 6, 80805 M\"unchen, Germany}}

\abstract{The goal of the T3B experiment is the measurement of the time structure of hadronic showers with nanosecond precision and high spatial resolution together with the CALICE hadron calorimeter prototypes, with a focus on the use of tungsten as absorber medium. The detector consists of a strip of 15 scintillator cells individually read out by silicon photomultipliers (SiPMs) and fast oscilloscopes with a PC-controlled data acquisition system. The data reconstruction uses an iterative subtraction technique which provides a determination of the arrival time of each photon on the light sensor with sub-nanosecond precision. The calibration is based on single photon-equivalent dark pulses constantly recorded during data taking, automatically eliminating the temperature dependence of the SiPM gain. In addition, a statistical correction for SiPM afterpulsing is demonstrated. To provide the tools for a comparison of T3B data with GEANT4 simulations, a digitization routine, which accounts for the detector response to energy deposits in the detector, has been implemented.
}

\begin{document}

\bibliographystyle{JHEP}

\section{Introduction}
\label{sec:Introduction}

Hadronic showers consist of a variety of different processes which lead to energy depositions in the active material of calorimeters, ranging from instantaneous signals from relativistic hadrons and  from electromagnetic cascades induced by the decay of neutral pions to delayed signals due to slow neutrons and nuclear de-excitation following neutron capture or spallation. This results in a complex time structure of hadronic showers in calorimeters, which depends on the choice of the active medium and of the absorber material. This may have consequences on the performance of detector systems in high-energy physics experiments, in particular in environments with high repetition rates and high background levels. 

Experiments at the planned Compact Linear Collider CLIC, a future  $e^+e^-$ collider operating at a center of mass energy of up to 3 TeV with a bunch crossing rate of $2\,\text{GHz}$ \cite{{Lebrun:2012hj}} are particularly challenging in this respect. The high luminosity and correspondingly high beamstrahlung, together with the high energy lead to high rates of $\gamma \gamma \rightarrow$ hadrons mini-jet events, which, in combination with the high bunch-crossing rate, results in significant pile-up of hadronic energy in the detector over a bunch train of 156 ns, far in excess of the energy deposited by a physics event. This beam-induced background can be substantially reduced by the use of particle flow event reconstruction \cite{Thomson:2009rp} with timing cuts on the nanosecond level in the calorimeter system \cite{Marshall:2012ry} in the highly granular calorimeters foreseen for linear collider detector systems \cite{Behnke:2013lya, Linssen:2012hp}. The evaluation of the performance of these algorithms in simulations based on GEANT4 \cite{Agostinelli:2002hh} requires an accurate modelling of the time structure of hadronic showers. The high collision energy at CLIC makes tungsten an attractive choice for the absorber medium of the barrel hadronic calorimeter for detector systems at CLIC, since its high density provides the necessary containment also for TeV-scale jets while allowing the calorimeter to be situated inside of the solenoidal magnet. 

As a heavy nucleus, tungsten is expected to have a richer time structure than stainless steel, due to the larger neutron component in the shower. Steel is considered as the absorber material for hadron calorimeters at the International Linear Collider ILC \cite{Behnke:2013xla}, which provides $e^+e^-$  collisions with energies of 500 GeV, and a possible upgrade to 1 TeV, putting less severe constraints on shower containment than the conditions at CLIC.  In addition to the expected increased importance of the time structure in tungsten, there is also very limited data on hadronic showers in tungsten to date. An experimental verification of the accuracy of the simulations of a tungsten-based calorimeter system is thus of importance for the development of calorimeter systems for future high-energy colliders. The {\it Tungsten Timing Test Beam} (T3B) experiment is designed specifically to provide a first step in this validation of the hadronic shower models. 

T3B was operated in test beams at the CERN PS and SPS together with the CALICE analog scintillator-tungsten hadron calorimeter (WAHCAL) \cite{LucaciTimoce201388} and, to provide reference data sets with steel absorbers, also with the CALICE semi-digital RPC-steel hadron calorimeter (SDHCAL) \cite{Laktineh:2011zz}. This experiment is an integral part of the CALICE program to study highly granular calorimeters. The analysis of hadron data, which will be the subject of two follow-up publications, provides radially and, for the case of tungsten, longitudinally resolved information of the average time structure observed with scintillator readout which are compared to simulations. In this paper, the technical implementation of the T3B detector is presented in detail in Section \ref{sec:Setup}, followed by a description of the data reconstruction including single photon timing, energy calibration and statistical afterpulsing correction in Section \ref{sec:Calibration}, and by a description of the detector simulation and simulated data digitization in Section \ref{sec:Simulation}.

\section{The T3B Setup}
\label{sec:Setup}

The T3B setup consists of two main components: The detector layer installed behind the CALICE hadron calorimeter prototypes, and the data acquisition system. 

\subsection{The Detector Hardware}
\label{sec:Setup:DetLayout}

\begin{figure}[t]
  \centering
  \includegraphics[width=0.99\linewidth]{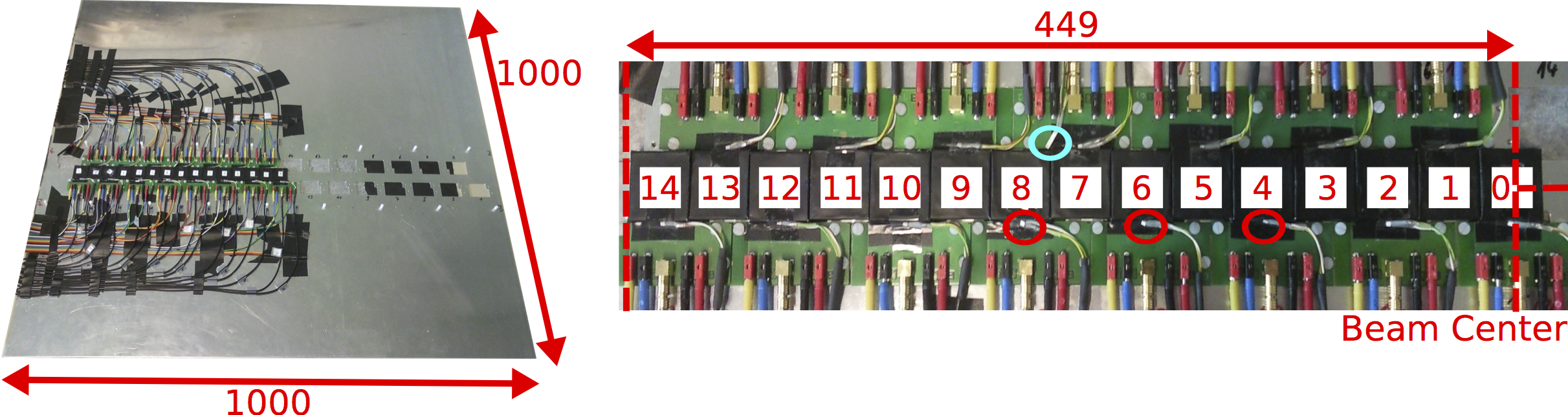} 
  \caption{Photograph of the whole T3B module (left) and zoomed view on the strip of scintillator
  cells (right). Each cell has its own temperature sensor that is attached close to the SiPM, shown by red circles for three example cells.
  In addition, on independent temperature sensor for the layer is installed,  marked with the blue circle. All dimensions are given in millimeters.}
  \label{fig:Setup:Setup}
\end{figure}

The T3B detector layer  consists of a strip of 15 scintillator cells mounted on an aluminium base plate with a thickness of $2\,\text{mm}$ as shown in 
Figure \ref{fig:Setup:Setup} (left) which is covered by an aluminium protection lid with a thickness of 1 mm. The lateral size of this cassette
structure is  $1\times1\ \text{m}^2$ with a total thickness of $13\,\text{mm}$, designed to fit into the mechanical structures of the CALICE calorimeter prototypes. The 15 scintillator cells are arranged in one horizontal strip extending from the center of the cassette out to a radius of 44.9 cm, as shown in Figure \ref{fig:Setup:Setup} (right), with a 1 mm tolerance gap between two cells. This arrangement provides full radial coverage of the hadronic shower. 

Each T3B cell consists of a scintillator tile machined from scintillating plastic (Bicron 420) with  dimensions of $3\times3\times0.5\ \text{cm}^3$. The size of the scintillator tiles matches the size of the fine scintillator cells used in the active layers of the WAHCAL \cite{Adloff:2010hb}. Each tile is read out by a Hamamatsu MPPC-50 with an active area of $1\times1\ \text{mm}^2$ and 400 microcells, packaged in a transparent plastic casing. The SiPM is coupled via an air gap without the usage of a wavelength shifting fiber to one side of the scintillator tile. To recover an uniform cell response to traversing ionizing particles, a special tile geometry is used with a dimple machined into the tile at the SiPM coupling position \cite{Simon:2010hf}. The tile with the embedded SiPM is completely enclosed with a self-adhesive highly reflective mirror foil (3M Radiant Mirror Foil) to provide optimal light collection, and a low-transmissive black absorber foil (BKF24 from Thorlabs) for light shielding purposes. The scintillator itself produces approximately 13\,000 photons per MeV of deposited energy. Only a small fraction of these photons, on the one to two per mille level, are actually detected, as presented in more detail in Section \ref{sec:Calibration:MIPcalib}.

Each cell is mounted on a custom designed preamplifier board based on an Infineon BGA614 amplifier, which amplifies the SiPM signal by a factor of 8.9 in
the relevant signal frequency range. The power for the 15 preamplifiers and the SiPMs is supplied by two common HV sources. Individually
adjustable resistor dividers allow for a device-by-device adjustment of the SiPM bias voltage prior to the installation of the detector
at the test beam facility. The operating voltages are adjusted such that the peak amplitude of the signal of a single firing pixel is 5 mV. To provide a detailed monitoring of the thermal conditions of the detector, each T3B cell is equipped with a four-wire temperature sensor (platinum resistance thermometer PT-1000) which is positioned close to the SiPM. In addition, an independent layer temperature sensor is installed in the T3B cassette, as shown in Figure \ref{fig:Setup:Setup} 
(right). All temperature sensors are read out with a PC via a custom USB-based temperature monitoring system.

\subsection{The Data Acquisition System}
\label{sec:Setup:DAQ}

The data acquisition of the T3B experiment is based on Picotech PS6403 USB oscilloscopes controlled with an external PC. These four-channel oscilloscopes provide a sampling rate of $1.25$ GSamples per second and 1 GB of local buffer memory. Two readout modes are available, a streaming mode where the recorded waveforms are transferred to the PC via the USB interface after each trigger, and a rapid block mode where a predefined number of events is recorded and stored in the local buffer memory, which are transferred after the number of events are reached. The acquisition of a data block can be stopped prior to reaching the full number of events by a signal from the PC, which starts the readout of all acquired waveforms. Since the streaming mode results in large dead-times between triggers which is incompatible with the need for very large data sets for the statistical analyses performed with the T3B data, the rapid block mode is used. The memory of the oscilloscopes is sufficient to record several 10\,000 events with a recording time of 2.4 $\mu$s before transferring the data to a PC for permanent storage. The oscilloscopes are triggered via an external trigger input, allowing synchronous operation of several devices. With the long recording window, trigger rates in excess of 100 kHz can be reached without event losses. The vertical resolution is 8 bits, with dynamic ranges of  $\pm50\,\text{mV}$ and $\pm200\,\text{mV}$ used for different acquisition modes explained below. Up to five such oscilloscopes are used in the experiment to record and digitize the SiPM signals of the 15 T3B cells and of supplementary trigger signals. 

The PS6403 oscilloscopes are controlled and read out via a PC running a C++ based data acquisition software with a QT-based graphical user interface specifically developed for the experiment. The software communicates with the oscilloscopes via the drivers provided in the software development kit of the manufacturer. 
The DAQ software provides a fully automated repetitive data acquisition loop that can be started and interrupted through the user interface and that can switch dynamically
between predefined sets of oscilloscope configuration settings based on external signals provided by the accelerator. The standard data acquisition loop is subdivided into a physics mode (PM) and an intermediate run mode (IRM), adapted to the beam structure at the CERN test beam facilities, with particles delivered in spills with a duration of a few 100 ms to 10 s, separated by periods without beam particles with a duration of a few to a few tens of seconds. 

In the PM, the T3B DAQ records the SiPM signals with the oscilloscopes triggered by an external beam trigger during a spill. After the end-of-spill signal is received, the DAQ stops the acquisition of data in the rapid block mode of the oscilloscopes and reads out the recorded data. In the IRM that follows immediately after the completion of the readout of the data collected during the spill, random thermal noise signals are recorded for each connected SiPM by directly triggering on the input channels with a threshold of 0.5 photon equivalents. These signals allow for a constantly updated determination of the SiPM gain and reference single photon signals, which are crucial for the signal reconstruction and calibration procedure of the T3B physics waveforms explained in the following section. The data recorded in the PM and in the IRM in one spill cycle of the accelerator are stored together as one spill of waveforms.  A physics run of the T3B experiment consists of many consecutively recorded spills, which are analyzed together as presented below.

\section{Signal Reconstruction and Calibration}
\label{sec:Calibration}

The signal reconstruction and calibration consists of several steps described in the following. First, individual p.e.\ signals are identified in the recorded waveforms. Then, depending on the goal of a particular analysis, a statistical correction for afterpulsing of the photon sensor is performed. Identified signals are calibrated to the energy scale given by the most probable energy loss of minimum-ionizing particles in a scintillator tile, and are corrected for temperature variations. Finally, timing corrections based on the signal amplitude and an overall trigger offset correction are applied.

\subsection{Signal Reconstruction Sequence and Waveform Decomposition}
\label{sec:Calibration:WfmDecomp}

\begin{figure}[t]
  \centering 
  \includegraphics[width=0.9\linewidth]{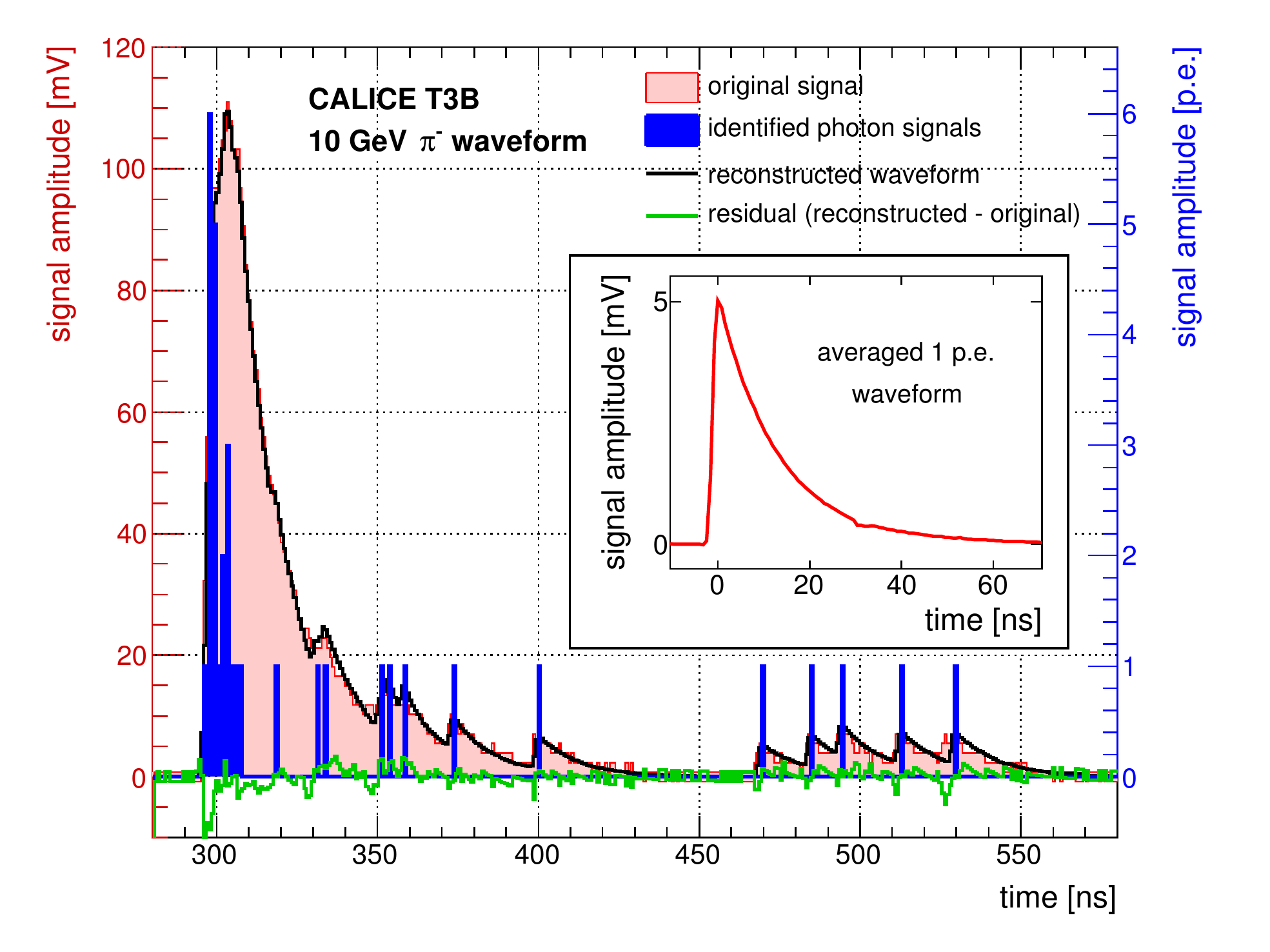} 
  \caption{Illustration of the waveform decomposition algorithm. The averaged waveform of a 1 p.e.\
  signal (plot inset) is subtracted iteratively from the analog physics waveform (red) resulting in a $1\ \text{p.e.}$ hit histogram
  (blue) representing the time of arrival of each photon on the light sensor. From this histogram, the original waveform is reconstructed back for cross
  checking purposes (black). The residual difference between the reconstructed and the original waveform (green) demonstrates the accuracy of the algorithm.}
  \label{fig:Calibration:WfmDecomposition}
\end{figure}

\begin{figure}[t]
  \centering
  \includegraphics[width=0.49\linewidth]{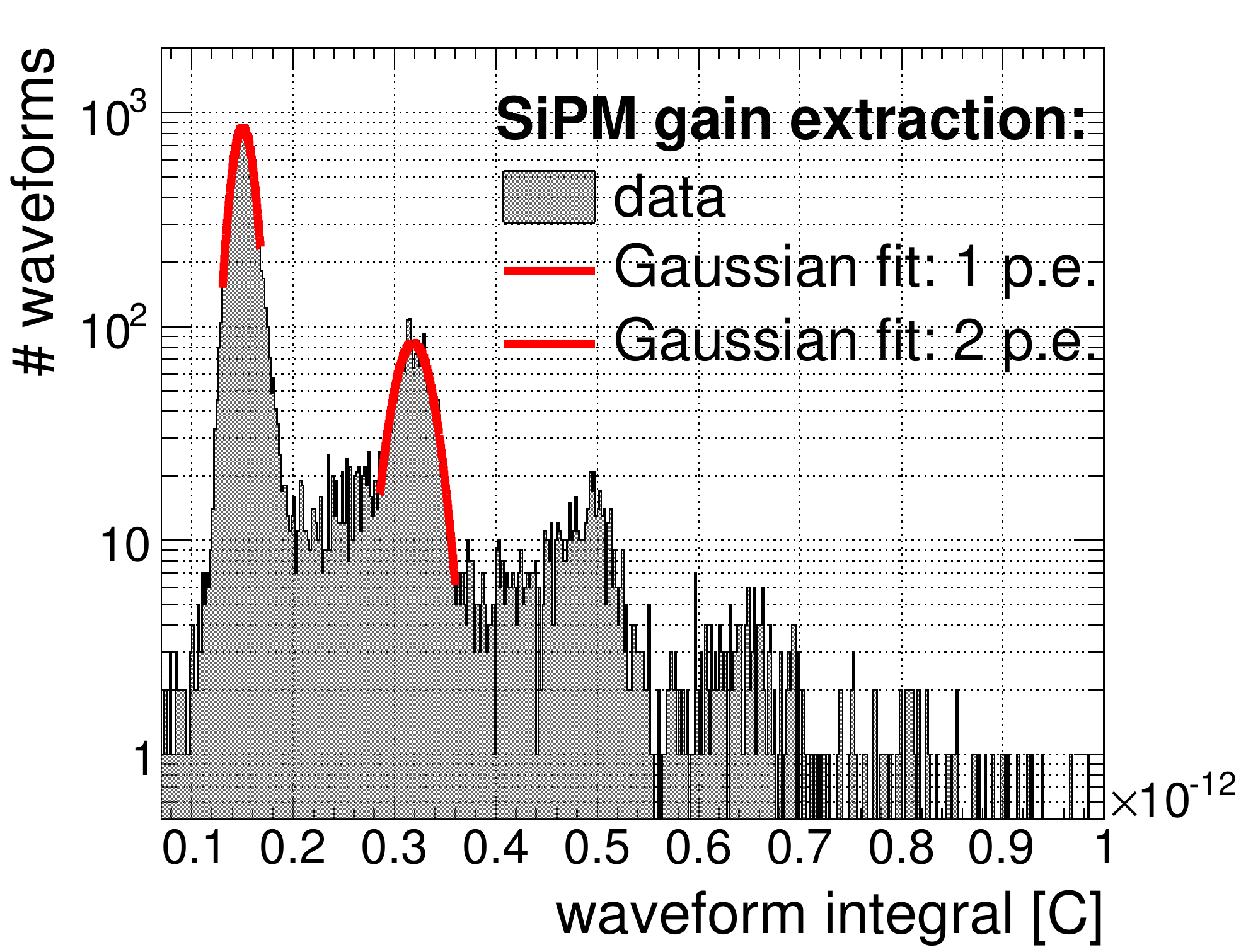}
  \includegraphics[width=0.49\linewidth]{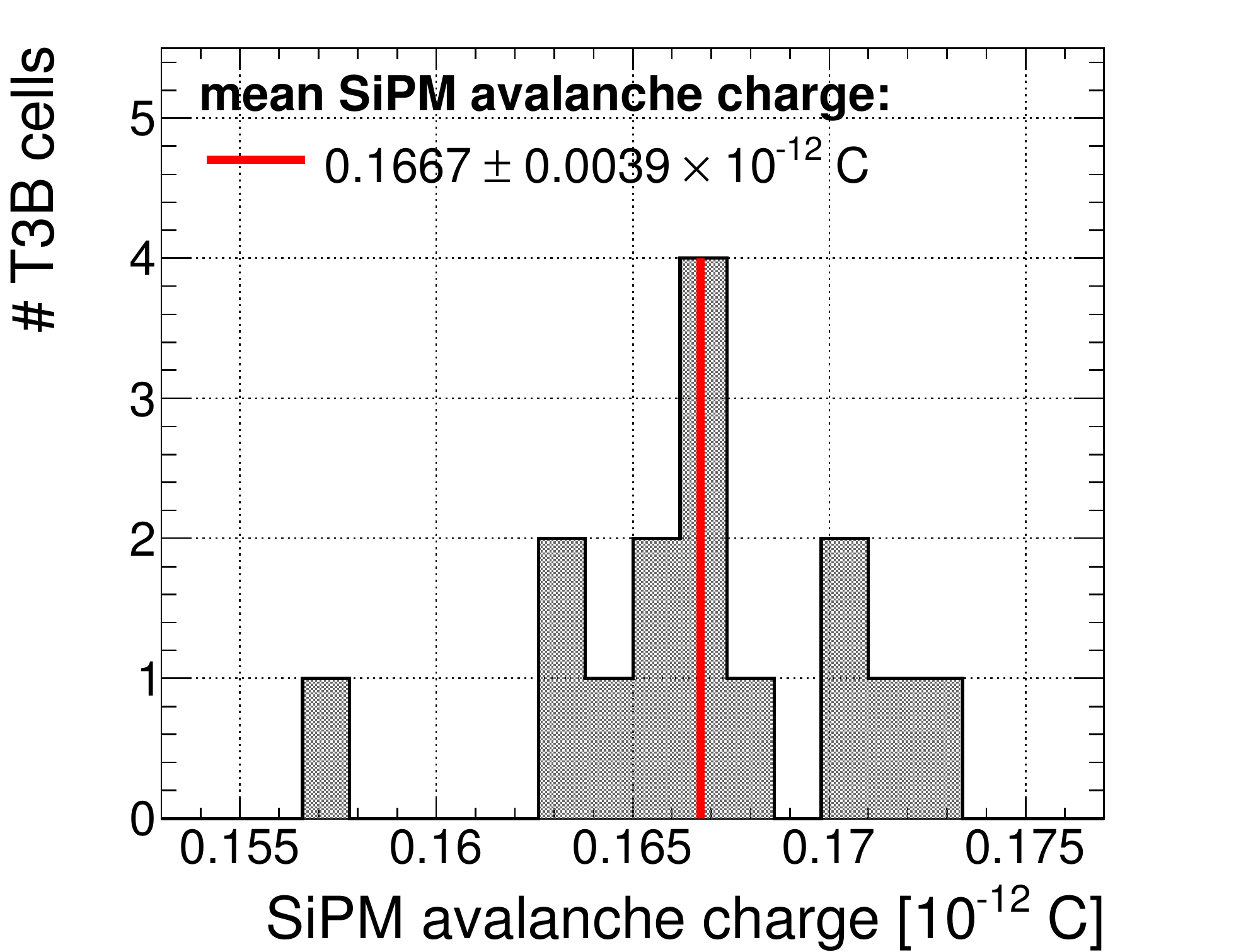} 
  \caption{Left: Extraction of the SiPM gain using  random thermal SiPM noise  acquired in the intermediate
  run mode. The SiPM avalanche charge value ($= \text{SiPM gain} \times e$) is extracted by fitting the 1- and 2-pixel peaks of the waveform
  integral distribution and by extracting the difference of the maxima. Right: Distribution of SiPM avalanche charge extracted from the fits of
  the waveform integral distribution for the batch of 15 T3B cells. The average charge of all 15 cells corresponds to a SiPM gain of $1.04 \times 10^6$.}
  \label{fig:Calibration:SiPMgain}
\end{figure}

The detection of a photon by one pixel of a silicon photomultiplier results in a characteristic and reproducible signal which is
characterized by a fast signal rise, related to the development of the electron avalanche, and slower signal decrease, related to the quenching
of the avalanche by the integrated quenching resistor. The rise time of the signal is approximately $2\ \text{ns}$ while the time constant of the exponential signal decay is
$12\ \text{ns}$ for the used MPPC-50 SiPMs coupled to the T3B readout system.

A waveform acquired in physics mode, e.g. the signal of the response of a T3B cell to ionizing particles, is an additive
combination of multiple 1-pixel signals occurring at different times, corresponding to the detection of scintillation photons. For tile traversing MIPs, which provide an instantaneous energy deposition in the scintillator, the time distribution of the recorded 1-pixel signals (also photon equivalents or p.e.) originates from the intrinsic time spread of the detected photons due to
scintillation time constants as well as from SiPM characteristics such as the occurrence of afterpulses and thermal darkrate. For data taken with hadron beams, additional late signal components arise due to delayed energy depositions in the hadronic cascade. The measurement of these contributions is the goal of the experiment. 

Figure \ref{fig:Calibration:WfmDecomposition} shows a representative waveform of  a hadronic shower event as measured by one T3B cell. The event was induced by a pion impinging with an
energy of $10\ \text{GeV}$ on a tungsten absorber structure. The waveform is characterized by a high main signal peak, followed by delayed signals of individual firing SiPM pixels.

Due to the temporal extent of 1 p.e.\ signals, their analog waveforms overlap considerably even for photon candidates detected several nanoseconds apart, as shown in Figure
\ref{fig:Calibration:WfmDecomposition} by the red histogram representing the original recorded waveform. As a consequence, it is non-trivial to distinguish the contribution of individual 1 p.e.\ signals and their time of detection. For a precise analysis of the time distribution of the T3B signals and for the quantification of energy depositions it is
desirable to determine the time of each photon detection with a precision on the nanosecond level. Thus, a waveform decomposition algorithm was developed which can extract this information from the recorded waveform. It is based on the iterative subtraction of a representative single 1 p.e.\ waveform from the recorded signal to determine the time of each photon candidate, and can, in principle, be applied to any SiPM signal where the full analog waveform is available. The use of a representative 1 p.e. waveform provides an automatic gain calibration of the procedure, eliminating the dependence of the data analysis on temperature-induced gain variations of the SiPM. Since the waveform decomposition transforms a full analog waveform with 3000 samples into a discrete set of photon equivalents with individual time stamps, it also provides a compression of the data volume by more than a factor of 100 without a loss of information. The waveform decomposition algorithm is implemented as a multi-step data reconstruction sequence, with initial calibration steps followed by the decomposition of the signal waveform. 

First, a pedestal subtraction is performed on a spill-by-spill basis for each T3B cell independently for data acquired in physics mode and in the intermediate run mode to eliminate possible pedestal shifts during data taking. From the pedestal determination, the noise of the full readout and DAQ system in physics mode is measured to be smaller than one LSB of the oscilloscope ADCs. Following the pedestal subtraction, the gain of each SiPM is determined from the IRM data. The pedestal-subtracted waveforms are integrated and histogrammed cell-by-cell, as shown in Figure \ref{fig:Calibration:SiPMgain} (left). The IRM contains primarily 1-pixel waveforms, but due to optical cross talk between SiPM pixels also two or more pixels may fire at the same time. This leads to a multi-peaked distribution in which each peak corresponds to a distinct number of fired SiPM pixels. From Gaussian fits to the first two peaks the gain is determined by the difference between the two maxima. For an accurate gain extraction at least 3000 waveforms are necessary. In normal beam operations this corresponds to a few spills (a few minutes).  In each measurement, the gain is determined with a precision of better than 0.5\%, given by the uncertainty of the determination of the peak-to-peak distance as shown in Figure \ref{fig:Calibration:SiPMgain} (left). Figure \ref{fig:Calibration:SiPMgain} (right) shows the distribution of the SiPM gain values for the set of 15 T3B cells, determined from the integral distribution of about 15.000 IRM waveforms per T3B cell. The relative standard deviation of the distribution $\sigma_{gain}/<gain>$, i.e. the spread of gain values around the mean, amounts to 2.3\%, showing a good degree of uniformity of the gain adjustment based on the resistor divider network. The mean signal amplitude of the MPPCs in nominal operating conditions is $1.67 \times 10^{-13}$ C, corresponding to a gain of $1.04 \times 10^{6}$. Following the gain determination, the average signal of a 1 p.e.\ waveform is determined for each cell from the IRM data. Based on the determined gain, single p.e.\ signals are selected by requiring waveform integrals matching the respective gain value within $\pm25\,\%$. The preselected waveforms are then averaged sample by sample. A typical resulting average waveform is shown in the inset in Figure \ref{fig:Calibration:WfmDecomposition}. The small dip observed at a time of 30 ns is due to an electronic reflection in the system. Since this reflection also occurs in physics data taking, it is included in the templates used for the decomposition algorithm. To obtain representative averages for the 1 p.e.\ waveform, at least 500 waveforms are required. Thus, also an averaged 1 p.e.\ waveform can be determined for every few minutes of data taking.

The waveform decomposition algorithm uses this averaged 1 p.e.\ waveform to determine the arrival time of each photon on the photon sensor. This is done by an iterative subtraction of this 1 p.e.\ signal from the physics waveform. First, the global maximum of the physics waveform is identified. The time position of the maximum of the 1 p.e.\ waveform is matched to the time bin of the global maximum, and the 1 p.e.\ waveform is subtracted from the physics waveform. The time position of the global maximum is taken as the time of the photon candidate. Then, the procedure is iteratively repeated for the resulting waveform after the subtraction until no significant signal remains. This is the case when no maximum higher than half of the peak amplitude of the 1 p.e. waveform is found, or the found maxima have a FWHM width which is smaller than 30\% of the FWHM width of the 1 p.e.\ waveform, which can occur from artefacts introduced during the decomposition procedure. These exit conditions were tuned to achieve a high efficiency for identifying all firing pixels, while avoiding the false identification of to many 1 p.e.\ signals. 
This waveform decomposition algorithm is well suited for an accurate determination of the time of arrival of photons on the light sensor.
The result of the waveform decomposition is a time resolved $1\ \text{p.e.}$ hit histogram which replaces the analog waveform, shown in blue in Figure
\ref{fig:Calibration:WfmDecomposition}. The accuracy of the waveform decomposition is verified by reconstructing an analog waveform from the  $1\ \text{p.e.}$ hit histogram by placing an averaged 1 p.e.\ signal at the time position of each identified photon candidate. The resulting waveform, shown in black in Figure \ref{fig:Calibration:WfmDecomposition}, is then compared to the original waveform shown by the red filled histogram. The accuracy of the waveform decomposition is demonstrated by the residual waveform (green), which shows the difference of the reconstructed and the original waveform. Negative residual values correspond to samples where not the full signal has been accounted for by the algorithm, while positive values correspond to too much subtraction. The resulting $1\ \text{p.e.}$ hit histogram is used for all analysis of the T3B data.

For test beam data, two different analysis approaches on the now reconstructed $1\,\text{p.e.}$ hit histogram are followed.
One analysis approach is to study the time distribution and the amount of energy deposited by the first hits of the T3B cells within an
event. Such first hits can occur promptly or significantly delayed within a hadronic shower event. A first hit is characterized through its
time of occurrence (time of first hit or TofH) and the amount of photon equivalents identified within a certain integration time window. To
identify first hits, a software trigger searches each waveform in an event for the signature of an energy deposition. For this identification, a threshold of 8 p.e.\ (equivalent to  approximately 0.5 MIP) and a trigger time window with a width of  $9.6\ \text{ns}$ is typically used. The
restriction of the trigger time window to only $9.6\ \text{ns}$ makes the analysis of first hits very robust against SiPM afterpulsing, as discussed in detail in Section \ref{sec:Calibration:Afterpulsing}. The actual number of found p.e., calibrated to the MIP energy scale as discussed in Section \ref{sec:Calibration:MIPcalib}, is assigned to the first hit as
its energy deposition. The time of the second identified p.e. is assigned as the time of occurrence of the first hit. Choosing the second
fired pixel reduces the probability that random thermal SiPM noise defines the timing of the first hit. The 1 p.e.\ noise levels of the SiPMs used in the setup are around 500 kHz, resulting in a probability of approximately 0.5\% for the occurrence of a thermal pulse in the trigger window before a real signal. Thus, taking the first identified p.e.\ as the time stamp would result in a slight deterioration of the time resolution of the system.

The second analysis approach uses the time of hit (ToH) of each identified photon.  To minimize the contribution from pure SiPM noise
waveforms, only waveforms in which more than $15\ \text{p.e.}$ (corresponding to $\sim0.6\,\text{MIP}$, since here the integration over the full time window is considered) were identified are considered in the ToH analysis. In terms of detected photon equivalents, the ToH analysis takes the full acquired information into account, including
delayed second hits of the same T3B cell within an event as well as the duration of cell hits (i.e. the time distribution of photon
equivalents within a cell hit) introduced by detector effects. On the other hand, afterpulsing plays an increasing role for later signals.  

Due to the small size of the T3B cells, multiple hits are quite rare even in highly energetic hadronic showers, and contribute only on the few percent level. Since the time of first hit analysis provides a superior robustness against detector effects, this approach is used for most analyses of the T3B data.

\subsection{Statistical SiPM Afterpulsing Correction}
\label{sec:Calibration:Afterpulsing}

\begin{figure}[t]
  \centering
  \includegraphics[width=0.99\linewidth]{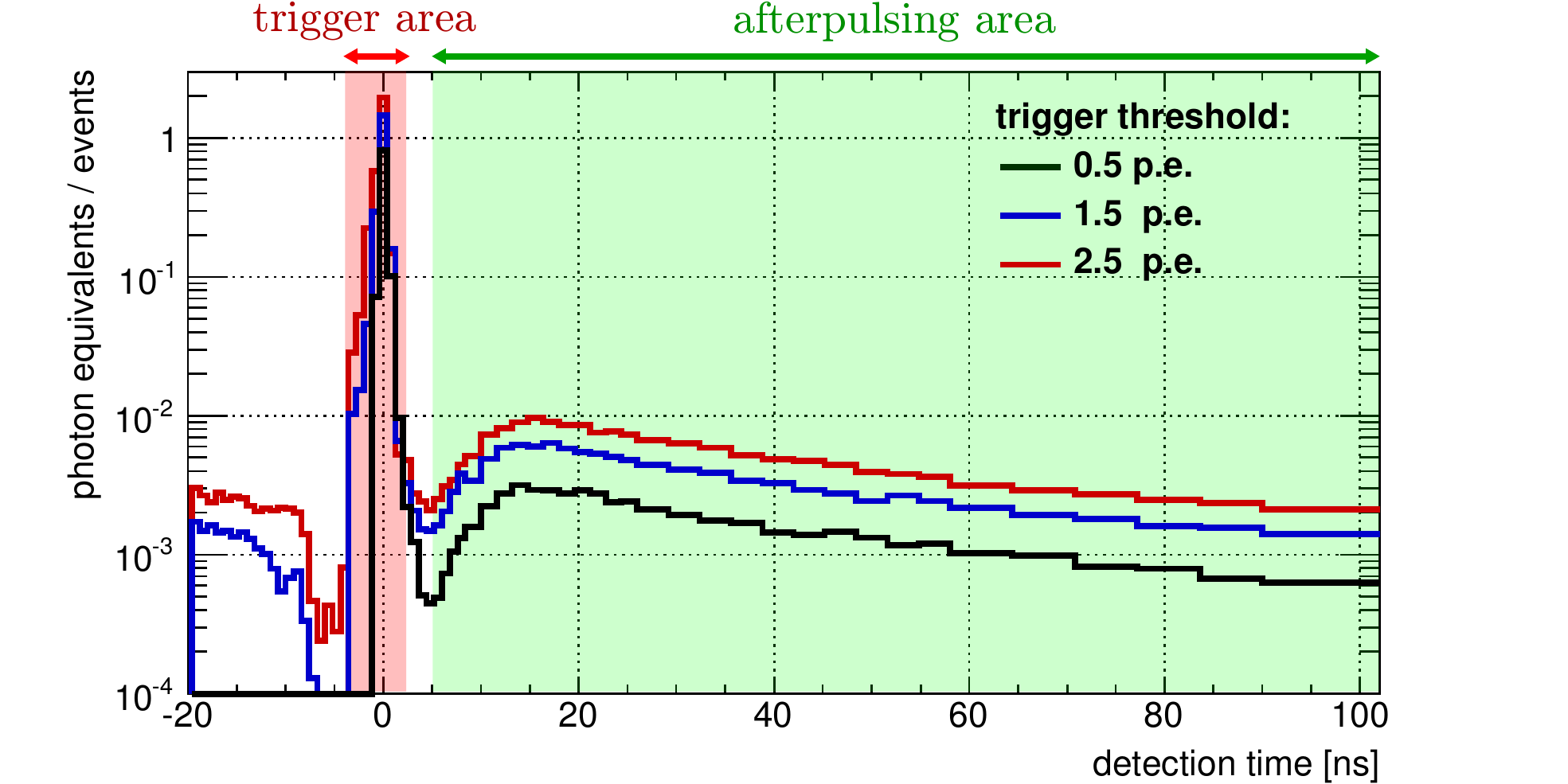} 
  \caption{Average time distribution of the afterpulsing of the central T3B cell for three different
  trigger thresholds. While the number of pixels firing in the trigger area (red) is fixed, the distribution of the pixels firing in the
  delayed afterpulsing area (green) is measured. The height of the AP distribution scales linearly with the number pixels contained in the
  trigger area.}
  \label{fig:Results:Afterpulsing}
\end{figure}

Afterpulses are generated by the delayed release of electrons trapped in the avalanche of a firing micro-cell, causing the same pixel to fire again, and thus result in additional pulses correlated with an initial signal. Since these afterpulses can occur with significant time delay, they have a strong impact on time distribution measurements in T3B when the full signal and not just the time of first hit is analyzed. As discussed in Section \ref{sec:Calibration:MIPcalib}, the interaction of minimum-ionizing particles results in typical prompt signals of about 17 p.e. Each of the fired pixels has a probability of approximately 20\% to afterpulse, resulting in significant late signals even without the presence of real delayed energy depositions. On an event-by-event basis, it is impossible to distinguish afterpulses from real photon signals. 
However, the contribution of SiPM afterpulses (AP) can be determined statistically, enabling the ToH analysis. Here, an average afterpulsing distribution is determined, which is then subtracted for every detected photon equivalent. 

The average afterpulsing distribution is determined from a series of laboratory measurements performed for different ambient temperatures and SiPM bias voltages to cover the range of parameters during the data taking at CERN. For the measurements, the T3B detector cells and the full DAQ chain are used. Temperature stability is guaranteed by a temperature controlled climate chamber. The oscilloscopes trigger on thermal dark pules of the
SiPMs. Three different trigger thresholds were used at $0.5\ \text{p.e.}$, $1.5\ \text{p.e.}$ and $2.5\ \text{p.e.}$, with one million events recorded for each setting.
The acquired waveforms are reconstructed using the waveform decomposition introduced above, and an event selection was applied which demands
exactly $1\ \text{p.e.}$, $2\ \text{p.e.}$ or $3\ \text{p.e.}$, respectively, within a  time window of $\pm2.4\,\text{ns}$ around
the trigger time, illustrated by the red area in Figure \ref{fig:Results:Afterpulsing}.

The contribution from pixels firing at a later point in time can be attributed to SiPM afterpulsing since the contribution from random
thermal noise is taken into account by a pedestal subtraction.  The increasing rate of afterpulsing for the first approximately 10 ns after the main pulse is due to the recovery of the fired pixels, which substantially reduces the probability for afterpulses while the bias voltage of the microcell has not been fully restored.  As expected, the afterpulsing distribution scales linearly with the number of fired pixels. Thus, the data taken with the three different trigger thresholds are combined after a division of the amplitudes by the number of p.e.
in the trigger area. The resulting time distribution is fitted with a function of the following form:
\begin{equation}
f(t) = \frac{erf(a \cdot t -b) + 1}{2} \times c \cdot (e^{-\frac{t}{\tau_1}} + e^{-\frac{t}{\tau_2}} + e^{-\frac{t}{\tau_3}}),
\label{eq:Results:AfterpulsingFitFunction}
\end{equation}
where $erf$ is the error function and a, b, c and $\tau_1$, $\tau_2$, $\tau_3$ are free parameters of the fit. This function describes the average distribution of afterpulses over a time window of 1 $\mu$s after a firing pixel. For each of the studied operating conditions (temperature and bias voltage) separate parameters are determined. The afterpulsing correction for a sample of events summed up in one histogram binned in time is implemented by iteratively subtracting the average distribution given by Equation \ref{eq:Results:AfterpulsingFitFunction}, scaled with the content of a given time bin, from all following time bins.

\begin{figure}[tb]
        \centering 
        \includegraphics[width=0.85\textwidth]{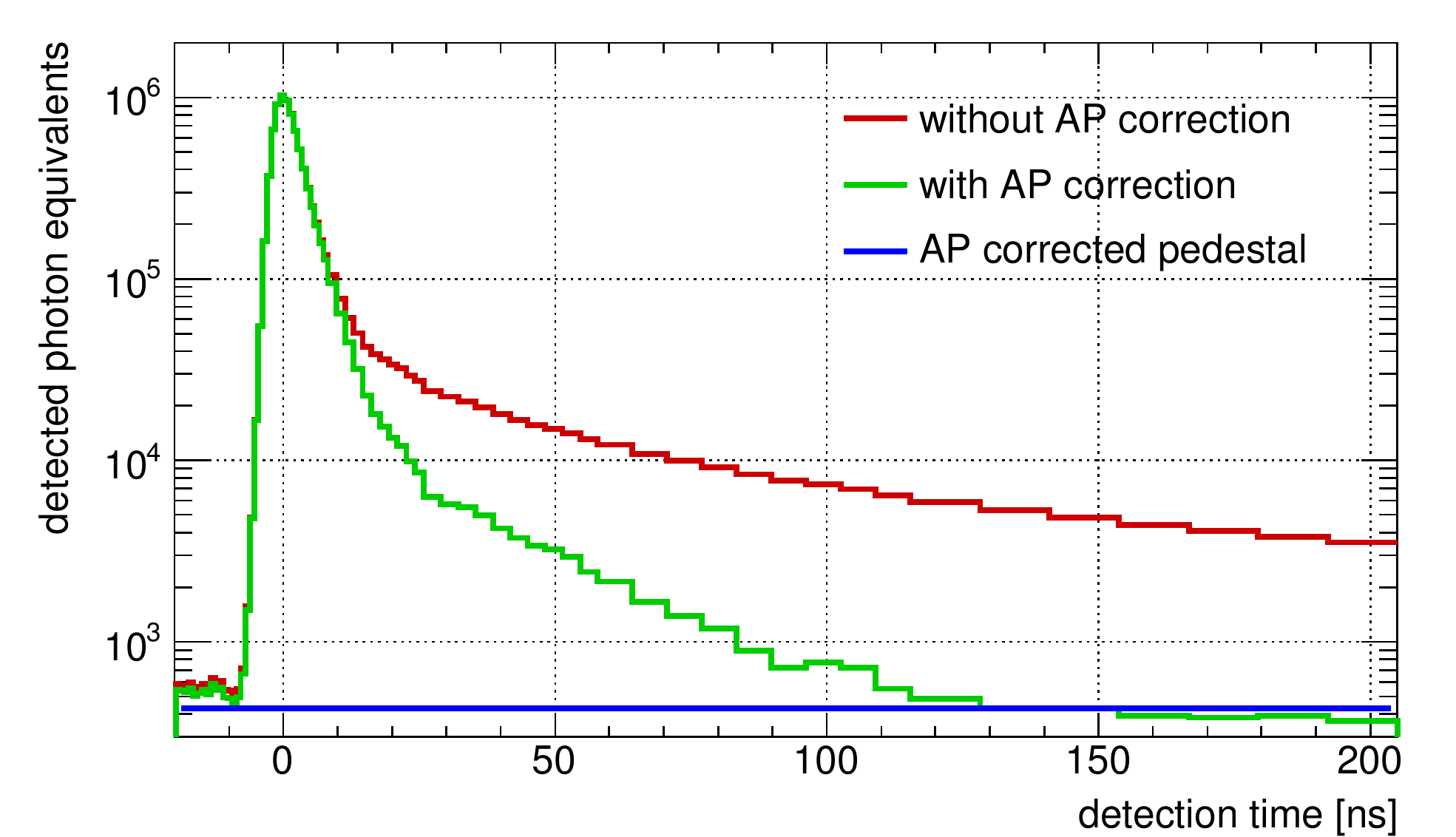}
        \caption[Afterpulsing Correction of the Time of Hit Distribution]{Time of hit distribution of the central T3B cell with (green) and without (red) applied afterpulsing correction for muon data at $180\,\text{GeV}$. The afterpulsing corrected pedestal value is shown in blue.}
        \label{fig:Results:APcorrection}
\end{figure}

Figure \ref{fig:Results:APcorrection} illustrates the effect of the afterpulsing correction, by showing the time of hit distribution of the central T3B cell for high-energy muons traversing the tile. These particles represent quasi-instantaneous energy depositions in the scintillator. The correction starts taking effect after approximately
$10\,\text{ns}$, the time when afterpulses set in after the recovery of the fired pixels.  The figure clearly shows that SiPM afterpulsing dominates the distribution at later times. The presence of signals after the correction at times later than $10\ \text{ns}$ is due to a long time constant of the scintillating material and a scintillation of the attached mirror foil. Nevertheless, at $50\ \text{ns}$ these contributions are nearly one order of
magnitude smaller than the contribution from afterpulsing and  almost vanish for times later than $100\ \text{ns}$. Here, the
distribution converges to the afterpulsing corrected pedestal which is determined in the pre-trigger range from $-110\ \text{ns}$ to
$-10\ \text{ns}$.

\subsection{Calibration to the MIP Energy Scale and Amplitude Temperature Correction}
\label{sec:Calibration:MIPcalib}

The purpose of the calibration to the MIP scale is to quantify individual energy depositions of hadronic showers in terms of the more
natural energy scale of the most probable energy loss of a minimum-ionizing particle in the scintillator cell, in short referred to as MIP. The number of detected photons for the energy deposition of one MIP depends on dynamical changes of the environmental conditions such as temperature changes which influence the photon detection
efficiency of a SiPM and on the details of the data reconstruction such as the time window over which an energy deposition is
integrated (and therefore implicitly on SiPM afterpulsing). These effects are corrected for in a calibration procedure which is derived from laboratory measurements and is applied to the data of each T3B cell during the analysis of hadron shower data.

\begin{figure}[t]
  \centering
  \includegraphics[width=0.49\linewidth]{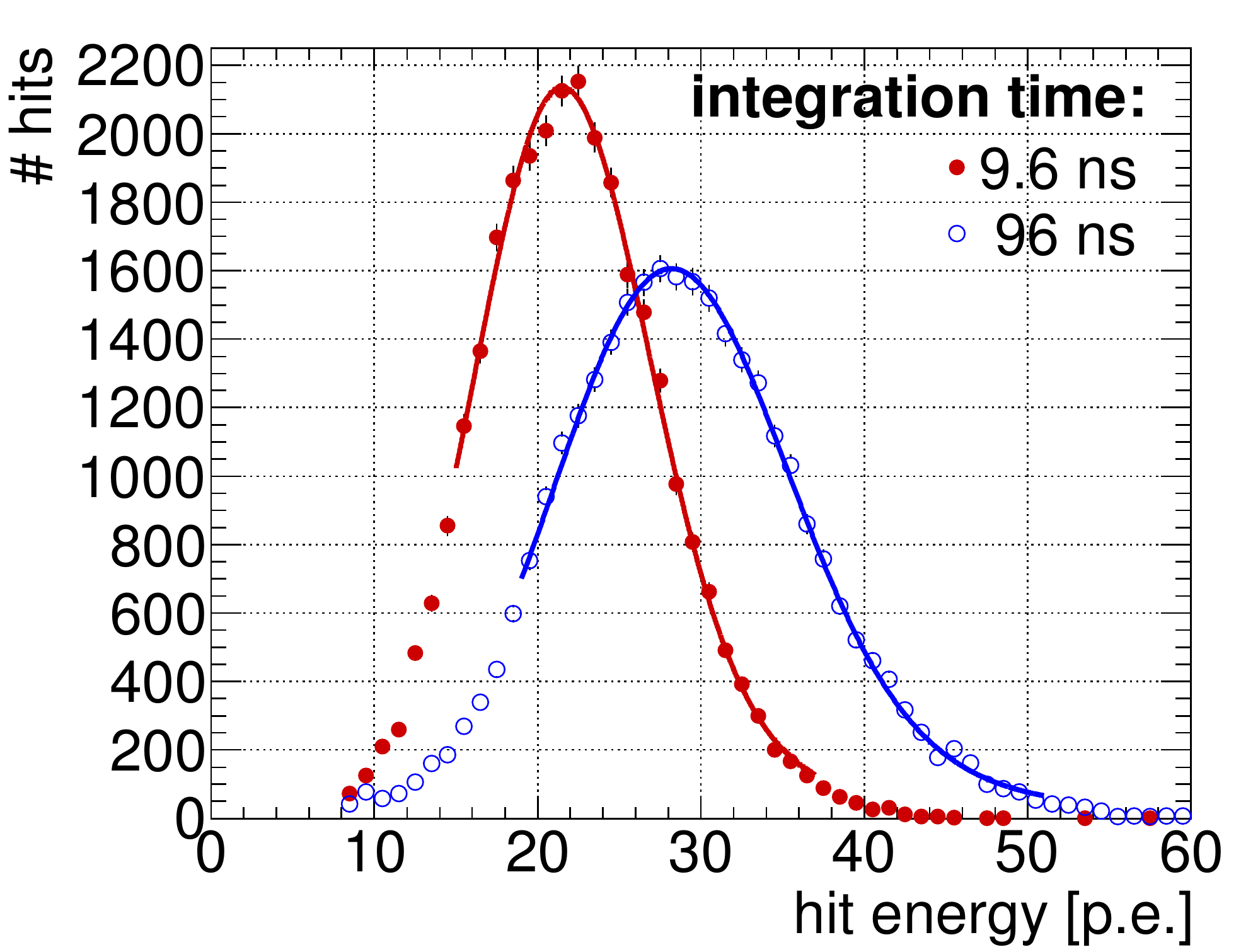}
  \includegraphics[width=0.49\linewidth]{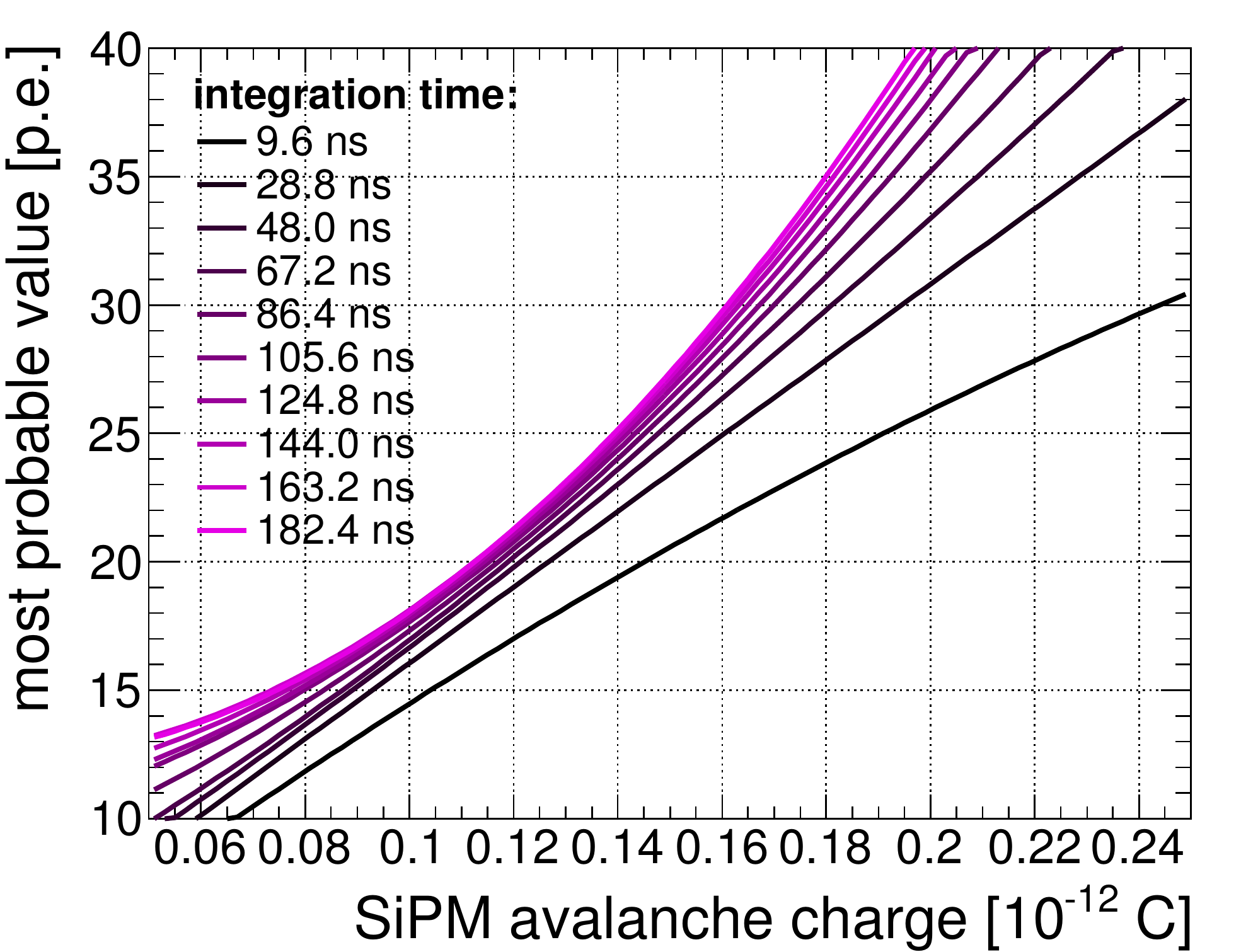} 
  \caption{Left: Distribution of the signal amplitude for $^{90}$Sr electrons for two different integration times. Due to the occurrence of afterpulses, the choice of the integration time strongly influences the most probable value of the signal for single particles. Right: Dependence of the MPV of the
  energy distribution on the SiPM avalanche charge ($= \text{SiPM gain} \times e$) for discrete time integration windows.}
  \label{fig:Results:Langau}
\end{figure}

The laboratory measurements are performed with a $^{90}$Sr source positioned above the T3B cell under study. A second T3B cell is located underneath, used as an additional trigger. The whole setup is located in a climate chamber to ensure a stable temperature during all measurements. The emitted electrons are collimated by a
tungsten casing with a circular opening of $1\ \text{mm}$ in diameter and point at the center of the tile under study.
The trigger settings of the T3B DAQ are adjusted such that a signal $>3\ \text{p.e.}$ has to be detected for both cells simultaneously.
This coincidence requirement ensures the selection of events in which the electron traverses the cell under study completely and thus
minimizes the acquisition of thermal SiPM noise. The signals of 20\,000 electrons are recorded and reconstructed by the waveform
decomposition routine explained above. Then, the energy distribution in terms of photon equivalents is determined for different time
integration windows in the range from  $9.6\ \text{ns}$ up to $182.4\ \text{ns}$. The obtained distributions are fitted with the convolution of a
Landau and a Gaussian function to extract the most probable value (MPV) of the distribution, as shown in Figure \ref{fig:Results:Langau} (left) for two different integration times.  The
increase in the extracted MPV for longer integration times originates primarily from the increasing number of afterpulses which are
integrated in addition to the photon signal generated by the tile-traversing particle.

This procedure is carried out for all 15 cells, and was performed for one reference cell for several bias voltage settings in the range of  $\pm500\ \text{mV}$ around the standard operating voltage. Since temperature changes result in a change of the breakdown voltage of the SiPM, the bias voltage scan is equivalent to a temperature scan. From the bias voltage scan, the dependence of the MPV on the SiPM gain is determined, and was found to be well described by a second order polynomial. Figure \ref{fig:Results:Langau} (right) shows this dependence, and illustrates the influence of the integration time. For short integration times, the dependence of the amplitude on the device gain is linear, while the increasing influence of afterpulsing for larger integration windows results in a parabolic behaviour.

The parameters determined from fits of the dependence of the amplitude on the gain are used to calibrate energy depositions to the MIP scale while eliminating the dependence on
gain variations at the same time. Due to the large temperature variations in a typical test beam environment this is a crucial step in the calibration of the data of the T3B experiment. This procedure makes use of the constantly updated gain values available during the analysis, as described in Section \ref{sec:Calibration:WfmDecomp}.

\begin{figure}[t]
  \centering \includegraphics[width=0.99\linewidth]{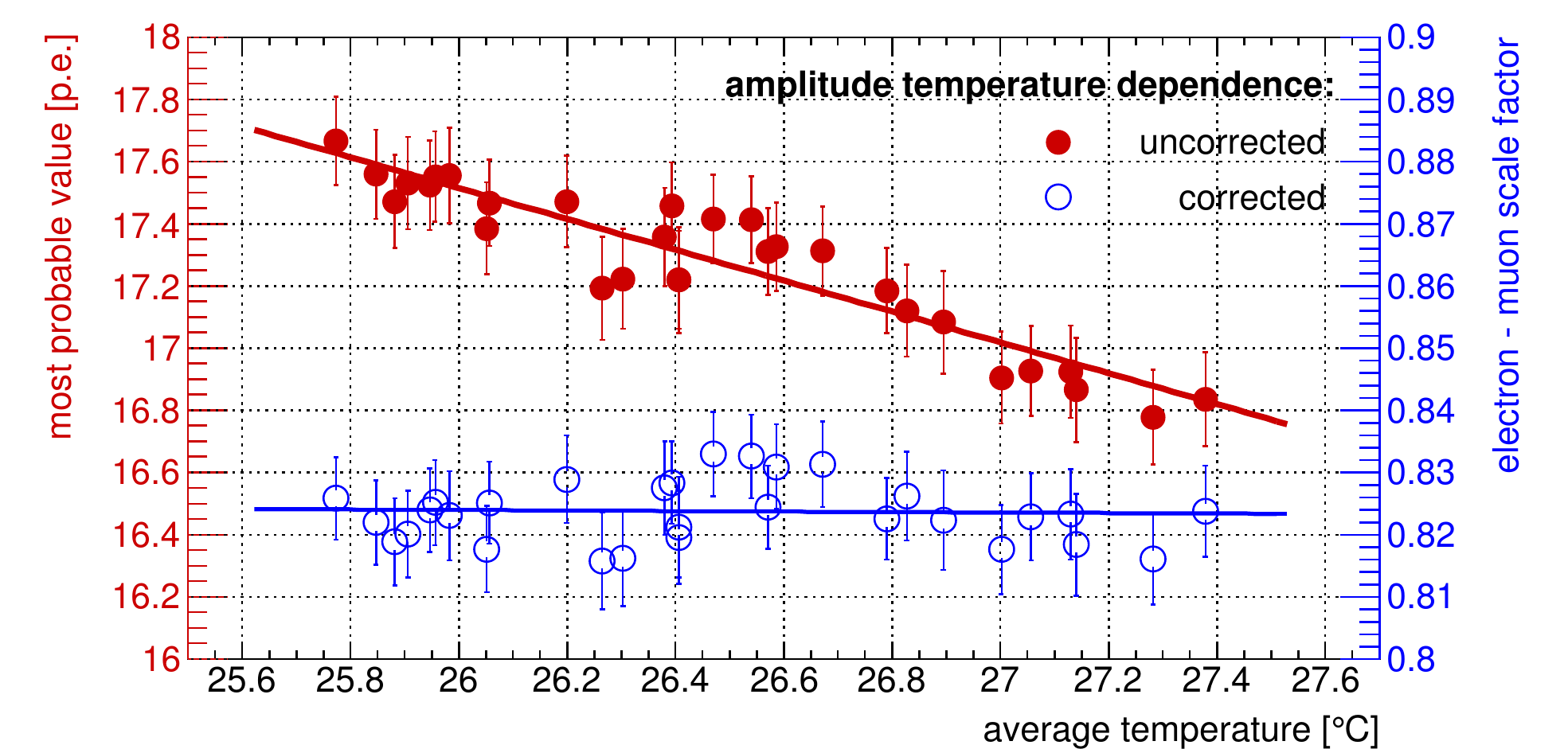}
  \caption{Validation of the calibration to the MIP scale with muon data for the central T3B
  cell and an integration window of $9.6\,\text{ns}$. The calibration procedure eliminates the temperature dependence of the MIP MPV and
  results in an average amplitude ratio between muons and electrons of 0.82, consistent with the difference in energy loss of electrons from the $^{90}$Sr source used in the laboratory to develop the calibration and of high-energy muons.}
  \label{fig:Results:MuonMPVTempDependency}
\end{figure}

A large muon data set recorded during one of the T3B test beam phases is used to validate the performance of this
calibration procedure. The data set consists of 13.4 million muon events that were acquired over a period of 40 hours without interruption. The data was split
into 30 sets in which the most probable value of the particle signals and the mean temperature was identified. The temperature varied in a range of $\sim2^\circ\text{C}$ due to
day-night temperature variations in the experimental hall. Figure \ref{fig:Results:MuonMPVTempDependency} shows the amplitude-temperature dependence of the central T3B cell
for the whole data set. It is characterized by an average MPV drop of $-3\,\%$ per Kelvin. Following the correction based on the laboratory measurements discussed above
for a time integration window of $9.6\,\text{ns}$, the temperature dependence of the signal amplitude is eliminated, as shown by the temperature corrected data points in Figure \ref{fig:Results:MuonMPVTempDependency}. Since the calibration procedure is developed on laboratory data using a $^{90}$Sr radioactive source, it determines the calibration factors relative to the most probable energy loss of an electron from the $^{90}$Sr  source which traverses the full tile. This energy loss, is higher than the energy loss of close-to-minimum-ionizing muons, thus the signal amplitude after calibration is given by an electron-to-muon scale factor $C_{e^- \Leftrightarrow \mu^-}$ of 0.82. This experimentally measured factor is consistent with predictions by GEANT4 simulations of the laboratory setup discussed in \cite{Simon:2010bd}, which yields a most probable energy loss of $^{90}$Sr electrons of approximately \mbox{980 keV} and of relativistic muons of approximately 805 keV. This scale factor is used throughout the analysis of the T3B data.

\subsection{Timing Corrections}
\label{sec:Calibration:TimeSlewAndTimingPrec}

To provide accurate measurements of the time of first hit, two corrections are required. The first corrects for a signal-amplitude dependent time slewing effect, and the second correction is a global time offset correction, which accounts for the a priori unknown offset between the time of the particle passage and the oscilloscope trigger, originating from signal processing and propagation times. This second correction, which is required to be able to perform comparisons between different data sets and between data and simulations, is also performed for analyses using all energy deposits.

\subsubsection{Time Slewing Correction}
\label{sec:Calibration:TimeSlewAndTimingPrec:TimeSlew}

\begin{figure}[t]
  \centering  
  \includegraphics[width=0.85\linewidth]{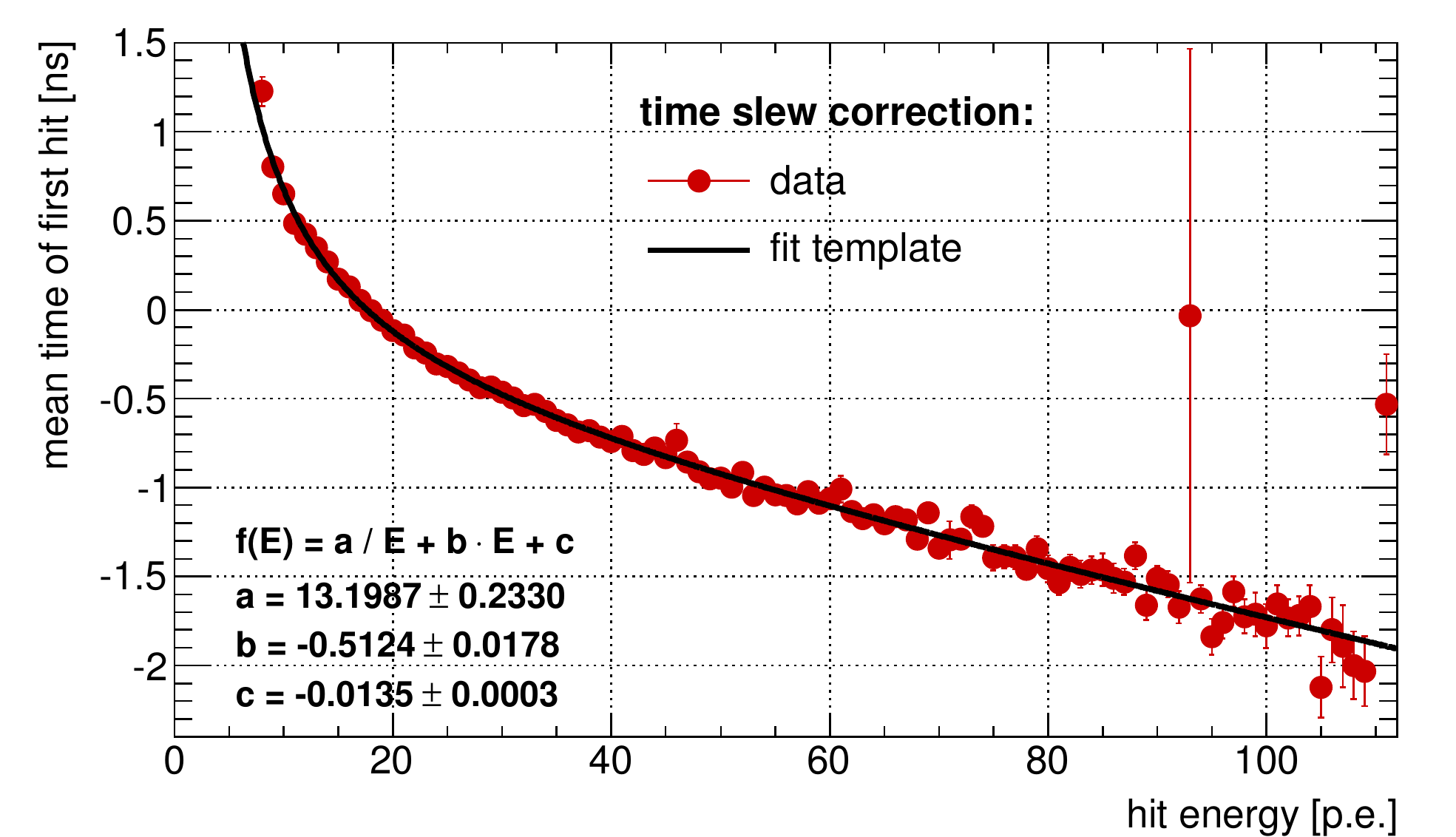} 
  \caption{Mean time of the first hit as a function of hit amplitude energy in the central T3B cell for 180 GeV muons. Zero time offset is defined by the timing of signals with an amplitude of 1 MIP. The dependence is fitted with Equation \protect\ref{eq:Results:TimeSlewCorrectionEquation}, which is used for the correction of the time slewing effect in the analysis.}
  \label{fig:Results:TimeSlewCorrection}
\end{figure} 

The use of a fixed threshold for the identification of the time of a hit in the detector results in a dependence of the determined time of a hit on the hit amplitude. For larger signals, the threshold is reached faster on average, resulting in earlier reconstructed times. To provide the highest possible accuracy for the time reconstruction over the full relevant energy range from 0.5 MIP up to several MIP, a correction for this time slewing effect is applied on a hit-by-hit basis. This correction is determined from muon data by studying the reconstructed time stamps of the signals relative to the beam trigger as a function of the signal amplitude, averaged over many events. Figure \ref{fig:Results:TimeSlewCorrection} shows the dependence of this mean time of first hit on the reconstructed energy, which was found to be well described by
\begin{equation}
\bar{t}_{Time Slew} = \frac{a}{E [p.e.]} + b \cdot E [p.e.] + c.
\label{eq:Results:TimeSlewCorrectionEquation}
\end{equation}

This function, with $a$ = 13.2, $b$ = -0.51 and $c$ = 0.0135, is used in the data analysis to correct the reconstructed time of each hit. Since the time slewing originates from the statistical distribution of the observed p.e., the correction is applied on the level of the uncalibrated hit energy prior to the MIP calibration.

\subsubsection{Time Offset Correction and Timing Precision of the T3B Cells}
\label{sec:Calibration:TimeSlewAndTimingPrec:TimingPrec}

Since the relative time difference between the signal of the passage of the particle and the trigger signal provided by the coincidence of several scintillators with further signal processing is a priori unknown, this offset has to be determined in the data analysis. This offset depends on the precise configuration of all systems used in the experiment and may change between different data-sets. The oscilloscopes used in T3B provide the possibility for extended pre-trigger recording, set to 400 ns during test beam operations, so even large offsets do not result in problems for the data acquisition. 

\begin{figure}[t]
  \centering
  \includegraphics[width=0.7\linewidth]{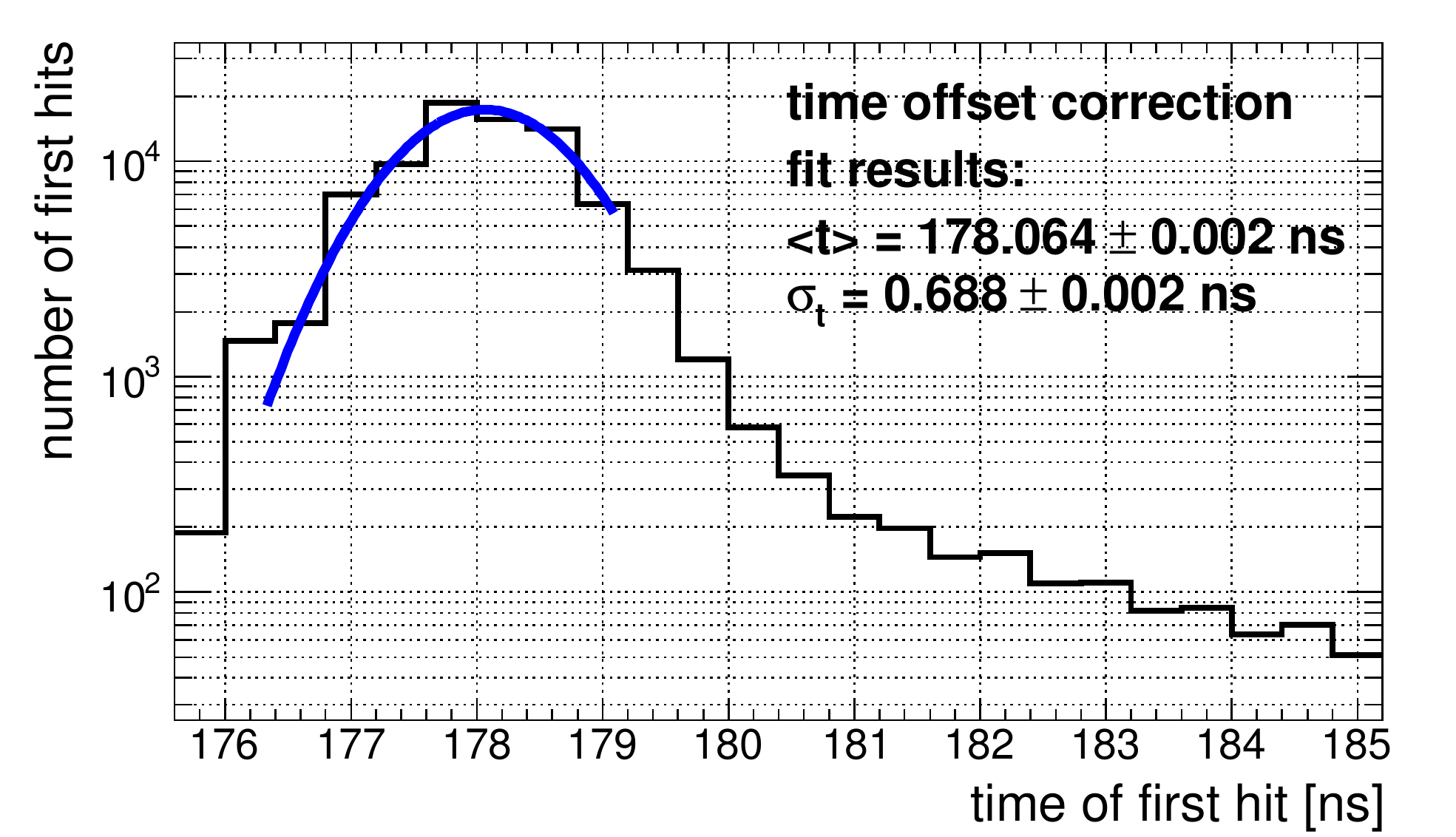} 
  \caption{Distribution of the time stamps for identified first hits in the central T3B cell for $60\,\text{GeV}$ hadrons taken together with the WAHCAL. The global time offset between the time of the particle signal and the oscilloscope trigger and the time resolution of the full system are determined from a Gaussian fit. Since the falling edge of the distribution at times $>178\,\text{ns}$ receives contribution from delayed components of the hadronic showers in addition to the intrinsic time resolution of the overall system the fit range is asymmetric.
  }
  \label{fig:Results:Timeres}
\end{figure}

Figure \ref{fig:Results:Timeres} shows the distribution of the time of first hit in the central T3B cell for a data set recorded with 60 GeV pions together with the WAHCAL.The main peak of the distribution represents the response of the  detector to the prompt energy depositions within showers, while the tail to larger times shows indications of later components of the hadronic cascade. Since the trigger configuration differs between muon and pion runs, muons, which provide only instantaneous energy depositions, can not be used to determine the time offset for hadron data. To extract the global time offset and the time resolution without substantial influence from the time structure of the hadronic cascade, the peak of the distribution is fitted with a Gaussian which only extends about 1 ns beyond the maximum of the distribution. From variations of the fit range and from the comparison of different data sets taken with identical trigger conditions the systematic uncertainty of the time offset determination, and with that the global systematic timing uncertainty of the experiment, is determined to be 200 ps. 

The overall time resolution of the system, which indicates the precision with which the time of the interaction of an individual particle in a T3B detector cell can be determined relative to the impact of the primary beam particle on the calorimeter, is given by the width of this prompt distribution, and depends on the intrinsic time resolution achievable with the T3B cells, the photon reconstruction and the time jitter introduced
by the used trigger configuration. In this sense, the width of the fitted Gaussian represents an upper limit of the TofH resolution achievable
with T3B cells. Depending on the specific configurations of a given test beam run, in particular on the layout of the beam trigger system, a time stamping precision of first hits in a range from
$0.7\,\text{ns}$ (for hadron data at $60\,\text{GeV}$ measured together with the CALICE tungsten analog calorimeter prototype) to
$1.5\,\text{ns}$ (for hadron data at $80\,\text{GeV}$ measured together with the steel semi-digital calorimeter prototype) was achieved with
the T3B experiment.

\section{T3B Simulations}
\label{sec:Simulation}

One of the main goals of the T3B experiment is the comparison of the recorded data to GEANT4 simulations with different hadronic physics models ("physics lists"). This requires an accurate implementation of the complete detector setup in the GEANT4 simulation framework, and a realistic modelling of the response of the T3B detector to the energy deposited by interacting particles, referred to as digitization. 

In the simulation, the T3B layers is implemented with two aluminium layers with a thickness of 2 mm and 1 mm, sandwiching the 5 mm thick scintillator tiles made of polystyrene and the printed circuit board of the preamplifiers modelled by a 1.7 mm thick layer of fiberglass. In addition, the CALICE calorimeters used in the different test beam phases of T3B, the WAHCAL and the SDHCAL, are implemented in the simulation setup. Data sets with protons, pions and muons with energies matching those of the data were generated with GEANT4 version 9.4p3 with the QGSP$\_$BERT, QGSP$\_$BERT$\_$HP and QBBC physics lists \cite{Geant4:PhysicsLists}, which differ in their treatment of neutrons in the hadronic cascade. 

To provide the information required for an accurate reproduction of the time structure of the detector response, the energy, location and time of each simulated energy deposition is saved during the event generation. Since low-energy neutrons, and with that corresponding recoil protons, play an important role in the late part of the cascade, a modelling of saturation effects in the scintillator for heavily ionizing particles, as described by Birk's law \cite{Birks1964}, is included in the simulation. 

From the generated events, the simulated detector signals are constructed with a dedicated digitization procedure which uses parameters derived from test beam data. This transforms the deposited energy given in units of eV to photon equivalents, and models the time response of the full T3B detector system. In a first step, the simulated energy depositions are assigned to individual detector cells based on their location. In each cell, the deposits are grouped in time bins of 0.8 ns, matching the sampling rate of the data acquisition. For each of these time bins, the energy content is rescaled from eV to p.e., which requires two separate calibration constants. The first is the transformation to MIP, which uses the most probable energy loss for relativistic muons in the scintillator tiles, which is determined from simulations to be  $C_{keV\leftrightarrow MIP} = 805.5\ \text{keV/MIP}$. The second is the transformation from the MIP energy scale to photon equivalents. The corresponding conversion constant $C_{MIP\leftrightarrow p.e.}$ is obtained from muon data, as discussed in Section \ref{sec:Calibration:MIPcalib}. Here, $C_{MIP\leftrightarrow p.e.} = 24\ \text{p.e./MIP}$ is obtained, since a  time window of 200 ns is used to account for photon sensor afterpulsing in the digitization. Since the number of detected photons per time bin is in general quite small, on the level of a few to a few ten photons, photon counting statistics are applied by randomizing the number of detected p.e. per bin from the simulated number based on a Poissonian distribution with the mean given by the integer number of p.e.\ obtained for a given time bin from the conversion from energy to detected photons. Finally, the time distribution of the detected photons for an instantaneous signal has to be accounted for in the simulations. This is done by randomly distributing the photons in one time bin of 0.8 ns according to a fit to the measured distribution of the signal of muons. The fit is obtained from the distribution without afterpulsing correction, shown in Figure \ref{fig:Results:APcorrection}, with a function of the form given in Equation \ref{eq:Results:AfterpulsingFitFunction}. After processing of the full information of a simulated event, this digitization procedure provides the time distribution of individual photon candidates in the same format as the output of the data reconstruction presented in Section \ref{sec:Calibration:WfmDecomp}. This allows the analysis of real and simulated data with the same analysis tools.

\section{Summary}

The T3B detector is a setup to study the time evolution of hadronic showers on a statistical basis in imaging sampling calorimeters with plastic scintillator active elements on the nanosecond level together with the CALICE calorimeter prototypes. It is based on a linear strip of 15 scintillator cells with a size of $3\times3\times0.5\,\text{cm}^3$ whose light signal is detected by silicon photomultiplier (SiPMs). The photon sensors are read out with commercial USB oscilloscopes with 1.25 GHz sampling and deep buffers to provide recording over several $\mu$s per event to sample the full evolution of the hadronic cascade. The experiment is read out and controlled with a custom C++ based data acquisition software run on a standard PC. 

For the reconstruction of the acquired data a specialized procedure was developed to obtain the time of detection of each individual photon from the scintillator. It provides an automatic gain calibration of the photon sensors by using single photon signals obtained from dark noise constantly during data taking. These reference signals are used to decompose the recorded waveforms into individual photon signals in an iterative procedure, providing the detection time for each photon with sub-nanosecond precision. 

The detector cells are characterized in the laboratory with a $^{90}$Sr source and with noise measurements to provide the necessary data for corrections for changing operating conditions such as bias voltage and temperature and for a statistical correction for afterpulsing. From hadron data taken at the CERN SPS, the overall timing capability of the system is determined, showing resolutions as good as 700 ps depending on the external beam trigger system. 

To provide the possibility for a comparison of data with GEANT4 simulations, a simulation framework with a detailed geometry implementation and a data digitization based on the detector response observed for muons, is developed. The analysis of hadron data, and the comparison to simulations with different hadronic shower models will be the topic of upcoming publications.

\section{Acknowledgements}

We thank the DESY FLC group for the MPPC photon sensors used in the experiment. The T3B project has been partially supported by the DFG Cluster of Excellence ``Origin and Structure of the Universe'' and by the European Commission under the FP7 Research Infrastructures project AIDA, grant agreement no.\ 262025.

\bibliography{bibliography} 

\end{document}